\documentclass[conference]{IEEEtran}

\usepackage{cite}
\usepackage{amsmath,amssymb,amsfonts}
\usepackage{algorithmic}
\usepackage{graphicx}
\usepackage{textcomp}
\usepackage{xcolor}
\usepackage{lipsum}
\usepackage{float}
\usepackage{booktabs}
\usepackage[per-mode=symbol]{siunitx}
\usepackage{caption}
\usepackage{subcaption}
\def\BibTeX{{\rm B\kern-.05em{\sc i\kern-.025em b}\kern-.08em
    T\kern-.1667em\lower.7ex\hbox{E}\kern-.125emX}}

\usepackage[left=0.625in,right=0.625in,top=0.75in,bottom=1.05in]{geometry}
\begin{document}

\title{Comparison of Tiny Machine Learning Techniques for Embedded Acoustic Emission Analysis}

\makeatletter
\newcommand{\linebreakand}{%
  \end{@IEEEauthorhalign}
  \hfill\mbox{}\par
  \mbox{}\hfill\begin{@IEEEauthorhalign}
}
\makeatother

\author{\IEEEauthorblockN{Uditha Muthumala, Yuxuan Zhang, Luciano Sebastian Martinez-Rau and Sebastian Bader}
\IEEEauthorblockA{Department of Computer and Electrical Engineering \\ Mid Sweden University, Sundsvall, Sweden \\ sebastian.bader@miun.se} 
}

\maketitle

\begin{abstract}
This paper compares machine learning approaches with different input data formats for the classification of acoustic emission (AE) signals. AE signals are a promising monitoring technique in many structural health monitoring applications. Machine learning has been demonstrated as an effective data analysis method, classifying different AE signals according to the damage mechanism they represent. These classifications can be performed based on the entire AE waveform or specific features that have been extracted from it. However, it is currently unknown which of these approaches is preferred. With the goal of model deployment on resource-constrained embedded Internet of Things (IoT) systems, this work evaluates and compares both approaches in terms of classification accuracy, memory requirement, processing time, and energy consumption. To accomplish this, features are extracted and carefully selected, neural network models are designed and optimized for each input data scenario, and the models are deployed on a low-power IoT node. The comparative analysis reveals that all models can achieve high classification accuracies of over 99\%, but that embedded feature extraction is computationally expensive. Consequently, models utilizing the raw AE signal as input have the fastest processing speed and thus the lowest energy consumption, which comes at the cost of a larger memory requirement.
\end{abstract}


\section{Introduction}\label{sec:introduction}

Acoustic emissions (AE) are a widely used non-destructive testing (NDT) method for identifying damages within concrete structures \cite{Xu2018}. The stress waves formed by the formation and propagation of cracks are the primary signs of potential structural failure. As a result, this technique is well suited for detecting such problems at an early stage. Researchers and engineers can identify, localize, and characterize the type of damage and provide correct measures to address the issue.

The damage mechanism in concrete involves microscopic strains during loading, leading to crack formation and propagation \cite{Carrasco2021}. These cracks release energy through stress waves, ranging from 10 kHz to 1 MHz \cite{Barbosh2021}. Characteristics of these AE signals, such as amplitude, frequency content, and rise time, are influenced by the damage type, loading conditions, and material properties \cite{Shahidan2017}. For example, tensile cracking generates higher-frequency AE signals than compressive cracking. These signals can be captured using piezoelectric transducers, which transform the elastic waves’ mechanical energy into electrical energy. The stethoscope-like structure of these sensors is thus able to capture the emissions from the structures correctly. This technology can be applied to concrete structures to detect defects during construction \cite{Bruncic2023}, provide in-service structural integrity assessment to identify deterioration \cite{Shahidan2017}, monitor bridges to ensure structural safety, and survey dams to detects leakage or structural problems \cite{Xu2018}. 

The captured signals contain essential information affected by the damage type. Figure \ref{fig:AE_waveform} illustrates an exemplar AE signal with several parameters that can be extracted from the waveform, representing the signal’s various characteristics, including amplitude, rise time, frequency, absolute energy, energy, signal strength, and central frequency \cite{Topolar2016, Shahidan2016}. Conventional approaches to AE damage classification rely on the manual interpretation of these waveform parameters \cite{LaMura2028} or the application of statistical thresholds. Manual interpretation is subjective to human expertise and prone to error, becoming impractical for complex structures due to time and expertise constraints \cite{Larosche2015}. Statistical thresholds may fail to capture the full range of damage types or their variability, leading to misclassifications \cite{Larosche2015}.

\begin{figure}
    \centering
    \includegraphics[width=.7\columnwidth]{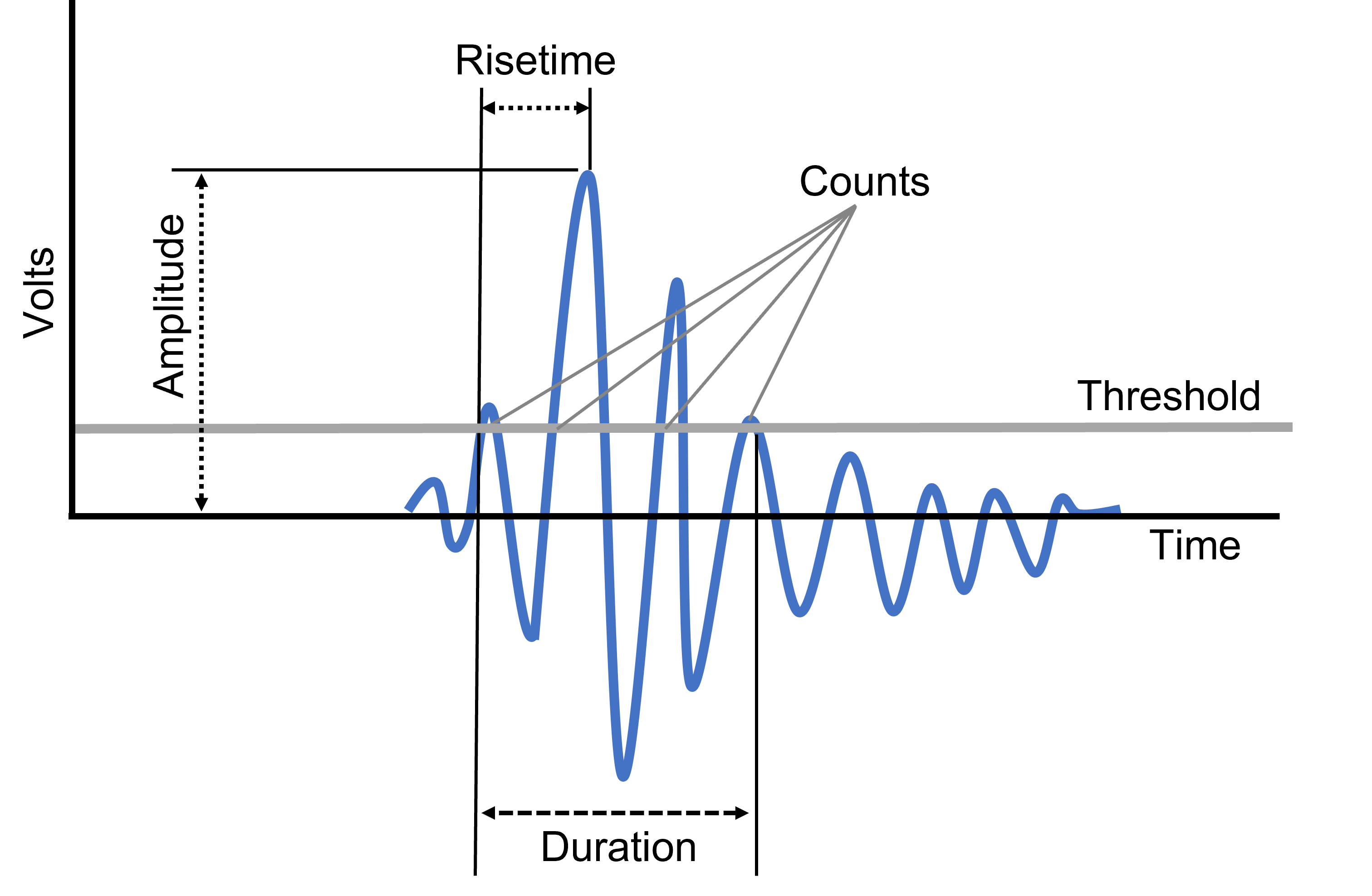}
    \caption{Overview of an example AE waveform with key properties.}
    \label{fig:AE_waveform}
\end{figure}

The advancement of machine learning (ML) has revolutionized AE signal analysis. ML models, such as neural networks, support vector machines, and decision trees, have demonstrated their ability to learn from large datasets of AE data, enabling accurate and efficient damage classification \cite{Thirumalaiselvi2021}. These ML models can handle complex signal patterns and noise, providing superior performance compared to traditional methods. However, implementing ML in AE analysis is not without challenges. Current ML models often have large sizes and computational requirements, making them impractical for real-time monitoring applications \cite{Zhang2022}. Additionally, integrating ML models into microcontrollers, commonly used in real-time structural health monitoring (SHM) systems, can be challenging \cite{Prem2017}.

Traditionally, a common approach was to use the extracted features of AE waveforms as the input to ML algorithms. However, more recently it has been demonstrated that ML models can extract the relevant information from the raw signal waveform \cite{Zhang2022, Zhang2023, Adin2023}. This paper aims to investigate which of the two approaches is preferable from an accuracy and energy consumption point of view when aiming for an IoT-based real-time implementation. While ML models such as neural networks will be significantly smaller and require less computation when operating on individual features rather than raw signals, the feature extraction itself will require extra computation and energy. Both approaches are deployed on a resource-constrained microcontroller, enabling real-time damage detection and classification. The models are evaluated in terms of accuracy, model size, inference time, number of parameters, memory, and energy per inference, demonstrating that raw-signal models have significant advantages.

The remainder of this paper is organized as follows. Section \ref{sec:method} describes the applied method and its implementation, including the utilized dataset, feature extraction and selection, as well as ML model design and deployment. Afterwards, Section \ref{sec:results} presents and discusses the obtained results, first for the models in general, and then with respect to their deployment. Finally, we conclude the study in Section \ref{sec:conclusions}.

\section{Methodology and Implementation}\label{sec:method}

The method in this article follows the typical design process of Tiny Machine Learning (TinyML) applications. It consists of feature extraction, feature selection, model design, training and validation, as well as deployment and testing of the generated model. The model architecture that has been used is a feed-forward Artificial Neural Network (ANN). As the target platform for the model, an Arduino Nano 33 BLE Sense was chosen, encompassing a low-power nRF52840 microcontroller with a ARM Cortex-M4 core. This microcontroller represents a standard-class microcontroller in many IoT applications.

\subsection{Acoustic Emission Dataset}

This paper utilizes a public dataset containing AE signals collected from a 15\,cm\textsuperscript{3} non-reinforced concrete block. The dataset is described in detail in \cite{Siracusano2021}. The dataset was created using a controlled experimental setup that employed five asymmetrically positioned piezoelectric AE sensors. AE signals were collected at a 5\,MHz sampling rate with 12-bit resolution, while the concrete block was stressed with a hydraulic press. The dataset contains 15\,000 AE events, each with a duration of 2 ms, initially comprising 10\,000 data points per signal. The signals reflect three distinct damage types, tensile, shear, and mixed, and are labeled accordingly. The dataset is thoroughly balanced, with an equal distribution of 5\,000 samples for each damage type, providing a comprehensive representation of the individual damage scenarios.

For classification, only measurements from the sensor with the highest signal-to-noise ratio have been selected. Each signal was down-sampled to 1\,000 data points to facilitate more efficient processing. Prior to further processing, the dataset has been partitioned with a 70\%, 15\%, and 15\% ratio for training, validation, and testing, respectively. This partition results in 10\,500 samples allocated for training and 2\,250 samples each for validation and testing.

\subsection{Feature Extraction}

Feature extraction is a common approach to reduce data dimensions of the ML model input. This process involves converting raw time-series data into essential features that effectively summarize critical information of the signals. The goal is to simplify the input data while keeping the most relevant information, aiding a more efficient classification. A comprehensive suite of 28 features has been identified for AE signal analysis, including statistical, temporal, frequency domain, and wavelet-domain features. These features are initially extracted using the Time Series Feature Extraction Library (TSFEL) in Python \cite{Barandas2020}, and are later implemented as C code for use on the low-power embedded system. The chosen features are designed to reflect the AE signals’ unique characteristics accurately.

Basic statistical features such as the mean, median, standard deviation, variance, maxima, minima, skewness, and kurtosis provide an overview of the signal’s central tendency, variability, and distribution shape. The mean and median relate to the average and center value of the signal, respectively, while standard deviation and variance measure the spread of signal values. Skewness and kurtosis give insights into the signal’s distribution asymmetry and tail heaviness.

Temporal features like absolute energy, zero crossing rate, auto-correlation, root mean square (RMS), peak-to-peak distance, slope, and counts of positive and negative turning points are extracted too. They capture the time-based dynamics of the signal. Absolute energy shows the signal’s total energy, RMS reflects its general magnitude, and zero crossing indicates how often the signal changes sign. Auto-correlation assesses the similarity between the signal and its delayed version, indicating patterns or periodicity. Peak-to-peak distance and slope provide information on amplitude changes and rate of change, respectively, while turning points help identify signal anomalies.

Frequency domain features, including the FFT mean coefficient, spectral entropy, fundamental frequency, spectral centroid, roll-off, and bandwidth, delve deeper into the signal’s frequency content. The FFT mean coefficient averages the Fourier Transform coefficients to highlight dominant frequencies. Spectral entropy evaluates the frequency distribution’s unpredictability, indicating signal complexity. The fundamental frequency, spectral centroid, roll-off, and bandwidth describe how energy is distributed across frequencies, shedding light on the signal’s harmonic structure and frequency range. A Fourier transformation is applied to convert the signal into the frequency domain before calculating the above features.

Wavelet features such as the wavelet absolute mean, energy, entropy, standard deviation, and variance have been shown to be useful for AE signals, which often show non-stationary behavior. These features allow multi-resolution analysis where the wavelet absolute mean and energy focus on the wavelet coefficients’ average magnitude and energy. Wavelet entropy is similar to spectral entropy but measures complexity or disorder in the wavelet domain. Wavelet standard deviation and variance indicate the spread of wavelet coefficients.

\subsection{Feature Selection}

Selecting the right features is a crucial step that significantly influences the system’s performance. The combination of input features will not only determine the classifier's accuracy, but the feature extraction will also require computational time and thus consume energy. For a resource-constrained embedded system, this means that we want to identify the smallest feature set that still provides sufficient classification accuracy.

This study initially employed the chi-square test and mutual information for feature selection, ranking features based on their predictive abilities. The chi-square test operates on the principle that a high score indicates a strong relationship with the target class, while mutual information measures the amount of information shared between a feature and the target class, with higher values suggesting a better predictive potential. However, applying these methods to the AE dataset provided inconsistent results as shown in Fig. \ref{fig:ranking}. Some parameters ranked high in one method, but low in the other (e.g., slope, median, kurtosis).

\begin{figure}[t]
    \begin{subfigure}{0.495\columnwidth}
        \centering
        \includegraphics[width=1\columnwidth]{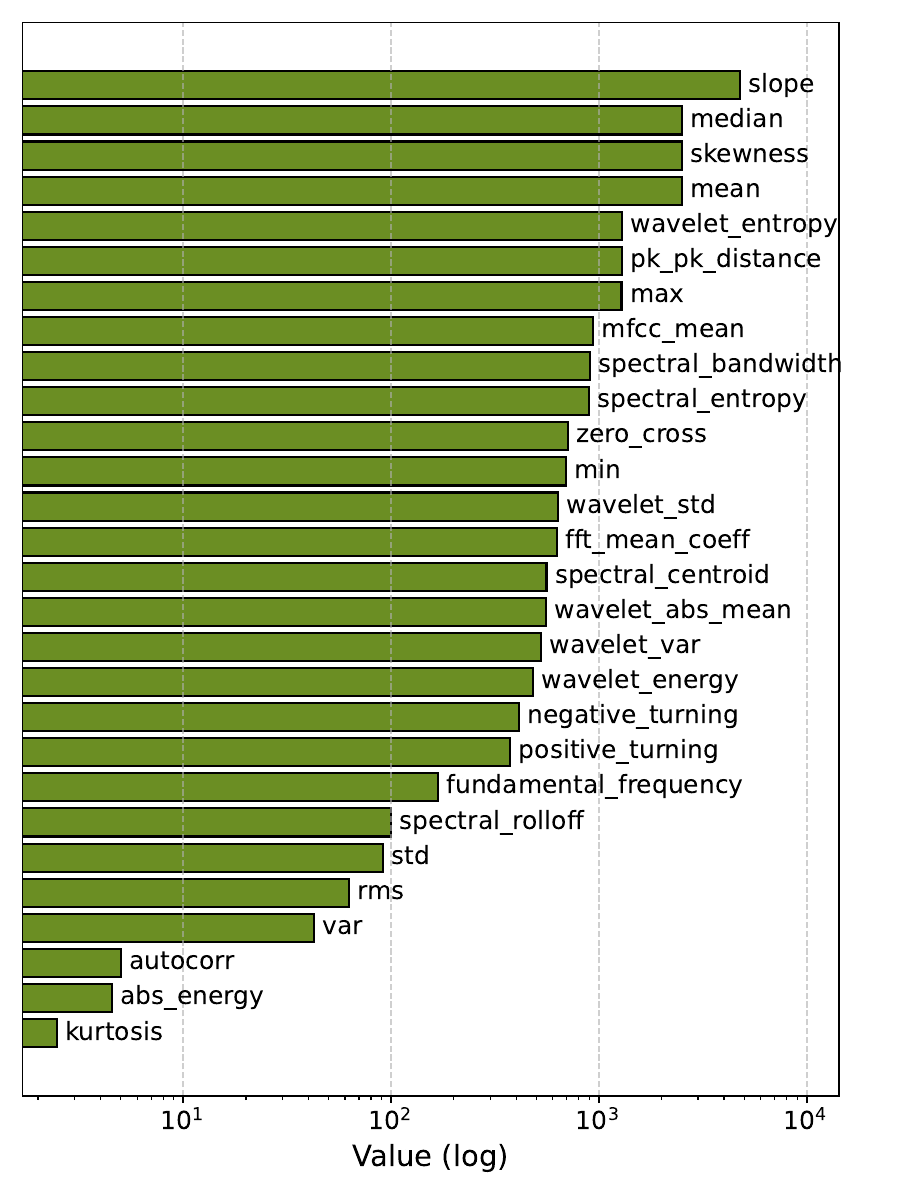}
        \caption{}
    \end{subfigure}
    \hfill
    \begin{subfigure}{0.495\columnwidth}
        \centering
        \includegraphics[width=1\columnwidth]{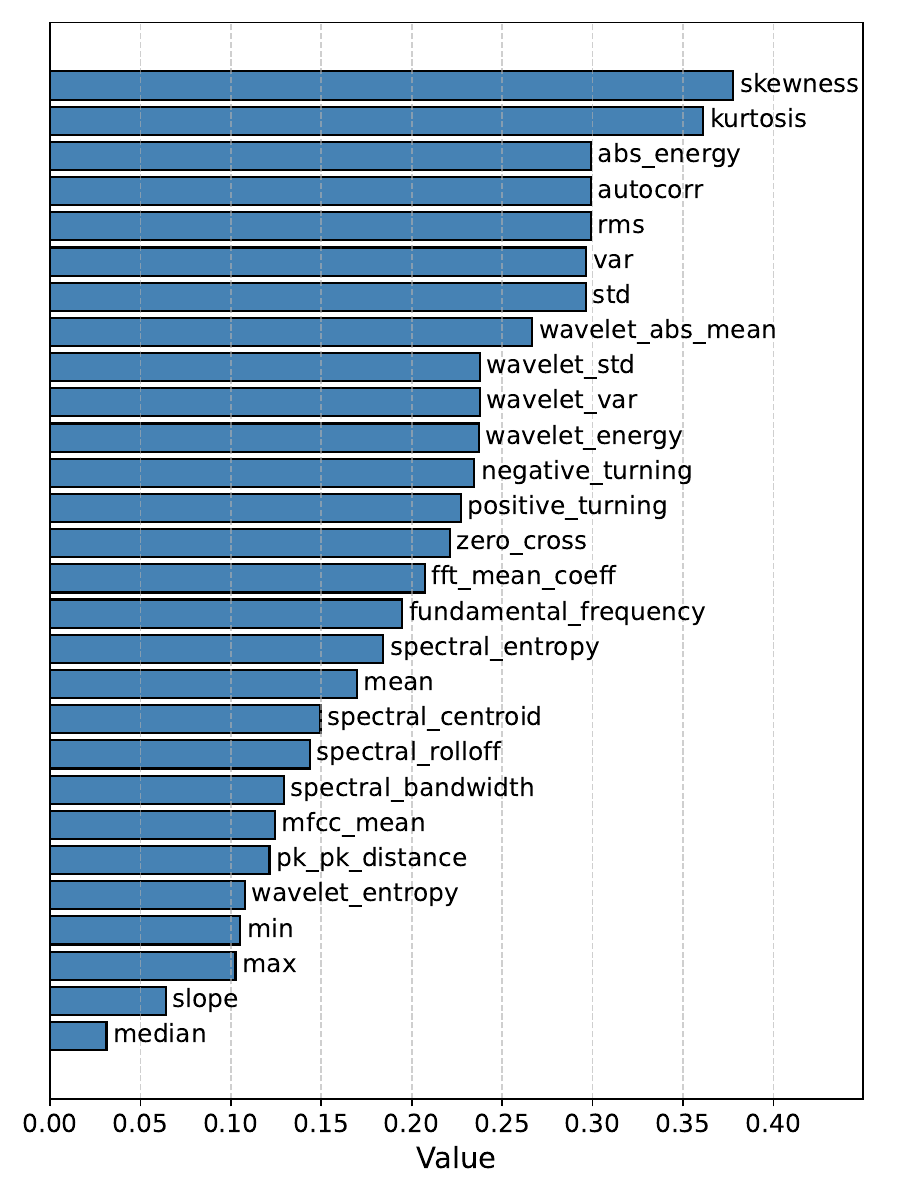}
        \caption{}
    \end{subfigure}
    \caption{Feature ranking according to (a) Chi-square test and (b) mutual information.}
    \label{fig:ranking}
\end{figure}

A more sophisticated method was adopted to address this issue, involving recursive feature elimination (RFE) with a decision tree algorithm. Decision trees are naturally suited for feature selection because they rely on information gained for making splits. The RFE process thoroughly examined every possible feature combination with a manually defined maximum set size of ten features. This analysis was conducted on the entire set of 28 features, but was also applied to specific sub-categories, such as only statistical and temporal features, assuming that frequency-domain features are more computationally expensive.

The feature selection approach suggested two suitable sets of input features. One set only contains time-domain (i.e., temporal and statistical) features, and is composed of 8 individual features: {'mean', 'standard deviation', 'skewness', 'kurtosis', 'zero-crossing rate', 'root mean square', 'peak-to-peak distance', and 'positive turning'}. The other set also contains frequency-domain features and is composed of 6 individual features: {'variance', 'kurtosis', 'peak-to-peak distance', 'negative turning', and 'FFT mean coefficient'}. 

\subsection{Artificial Neural Network Model Design}

The design of the ANNs has been tailored to the unique characteristics of input data to provide the best performance for each input condition. As a result, three distinct ANNs have been implemented and optimized, namely one for the raw AE signal input (raw-signal model), one for the 8 time-domain features (8-feature model), and one for the 6 time- and frequency-domain features (6-feature model). The raw-signal model uses a 1000x1 input vector, containing all samples of the down-sampled signal. As such, for this model no prior feature extraction is necessary. The architectures of all three ANNs are depicted in Fig. \ref{fig:ann_architectures}. Each model is comprised of an input layer, two hidden layers, and an output layer. The input layer size is defined by the input data, and the output layer size is three (i.e., according to the three damage types). The number of neurons in the hidden layers is optimized for each model respectively. Moreover, the hidden layers use ReLU activation functions, whereas the output layer uses SoftMax. The models are implemented in Python using TensorFlow with the Keras API.

\begin{figure}[t]
    \centering
    \includegraphics[width=\columnwidth]{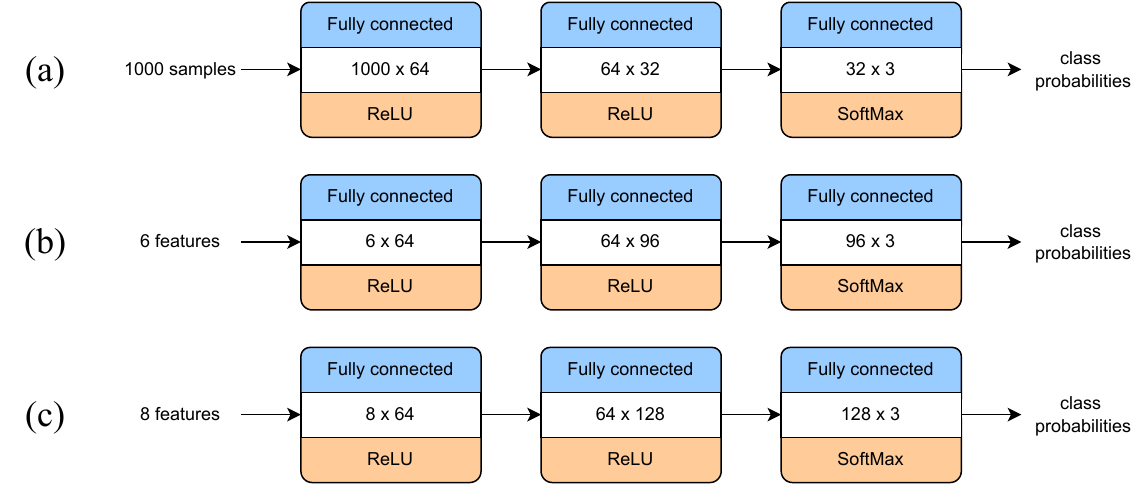}
    \caption{Architectures of the three ANN models for different input data (a) raw AE signals, (b) 6 time- and frequency-domain features, and (c) 8 time-domain features.}
    \label{fig:ann_architectures}
\end{figure}

\begin{figure*}
    \begin{subfigure}{0.3\textwidth}
        \centering
        \includegraphics[width=\columnwidth]{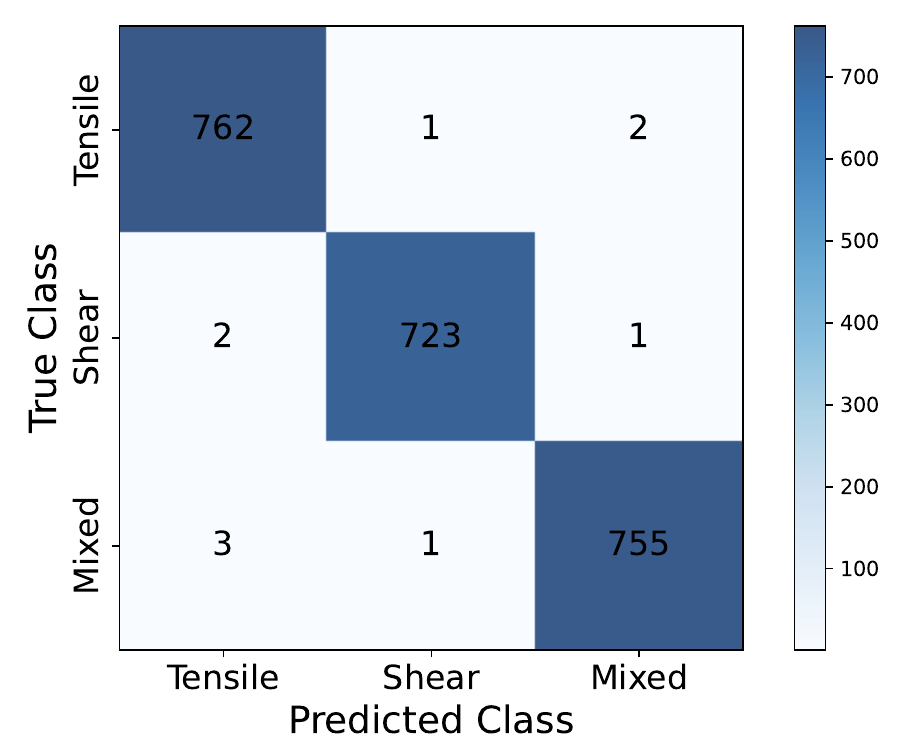}
        \caption{}
    \end{subfigure}
    \hfill
    \begin{subfigure}{0.3\textwidth}
        \centering
        \includegraphics[width=\columnwidth]{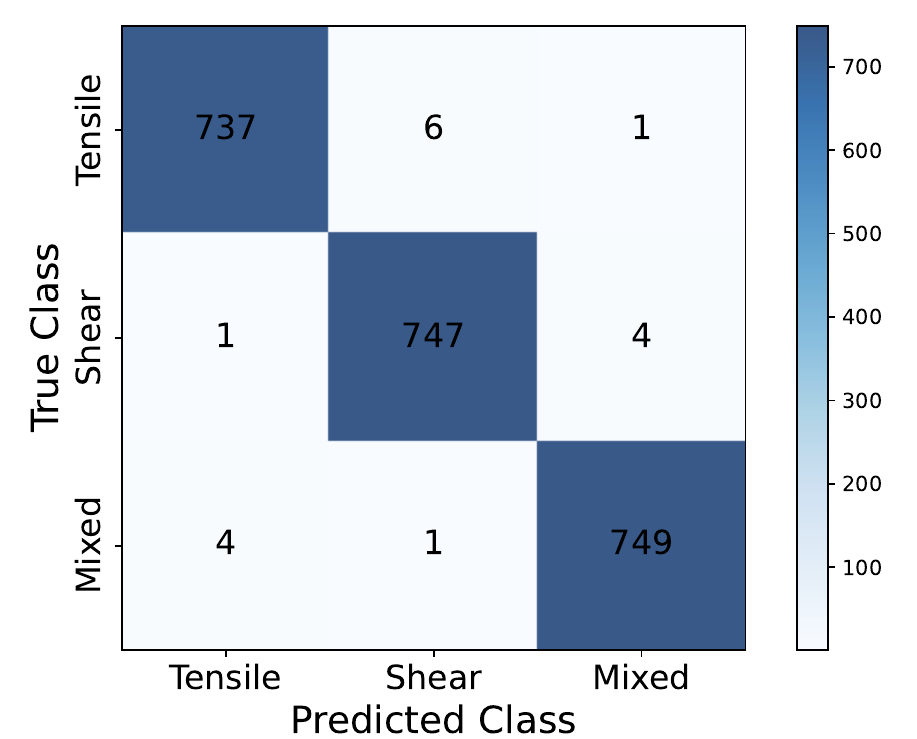}
        \caption{}
    \end{subfigure}
    \hfill
    \begin{subfigure}{0.3\textwidth}
        \centering
        \includegraphics[width=\columnwidth]{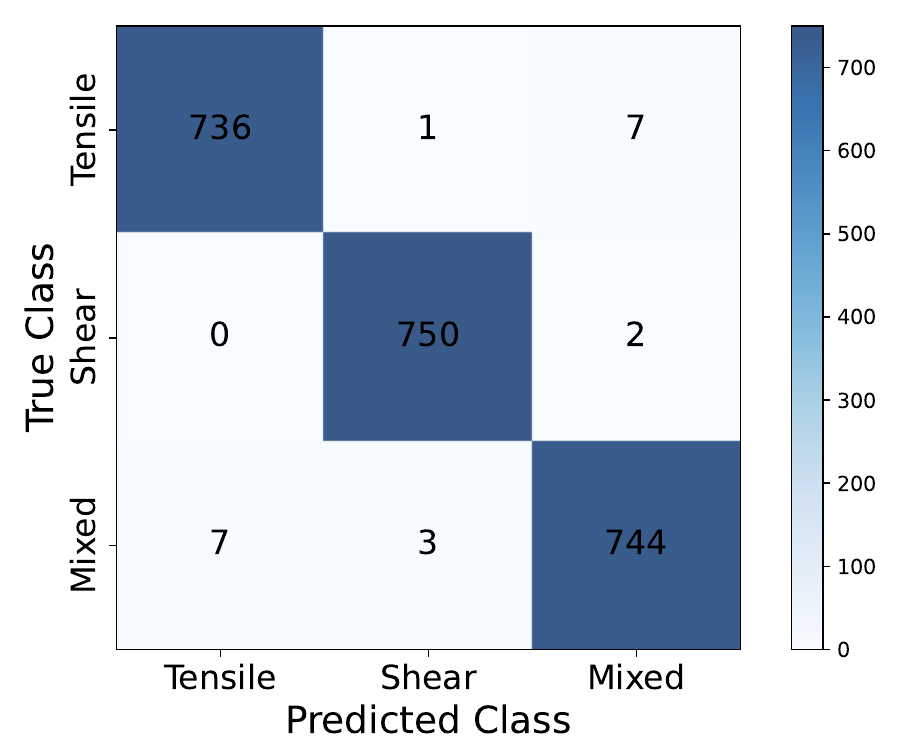}
        \caption{} 
    \end{subfigure}
    \caption{Confusion matrices of test set evaluation for the (a) raw-signal model, (b) 6-feature model, and (c) 8-feature model.}
    \label{fig:confusion_matrices}
\end{figure*}

Keras Tuner is used to automate the optimization of hyperparameters for each model \cite{Pon2021}. Keras Tuner systematically explores various combinations of neurons in each layer, as well as the learning rates, thereby determining the most effective model parameters through an objective performance metric. The three-layered structure standard for all models with two hidden layers was designed to balance model complexity and computational efficiency. The ReLU activation function in the hidden layers is mathematically represented as f(x) = max(0,x), which activates a neuron only if the input is positive, thus introducing non-linearity without affecting the sparsity of the model. The output layer’s SoftMax function provides a probabilistic interpretation of multi-class predictions beneficial for categorical outcomes. 

Finally, the optimum number of epochs for training was achieved using an early stopping mechanism. Early stopping is a form of regularization used to avoid overfitting by terminating the training process if the model’s performance on a validation set does not improve for a specified number of epochs. This strategy ensures that the model’s learning is halted when sufficiently generalized, thereby preserving its predictive performance on unseen data. Before deployment, the models were evaluated based on the parameter number, accuracy, and model size.

\subsection{Model Deployment}

With feature selection and model design in place, they need to be deployed on the low-power microcontroller for efficient operation. Custom implementations for the extraction of each selected feature was implemented in optimized C-code. Each function was meticulously evaluated against the initial Python code to verify equivalent results. For FFT implementation, the Arduino FFT library was used.

Afterwards, the developed ANN models were optimized for microcontroller deployment. The models were converted using the TensorFlow Lite for microcontroller framework (TFLM), applying int8 quantization to model parameters. TensorFlow Lite converts models into a format suitable for resource-constrained devices while maintaining adequate accuracy for inference tasks. Only required resolvers are included in the code to minimize the memory overhead of the TFLM framework. Each model and its selected features were then individually tested on the microcontroller to evaluate its real-time performance. This testing procedure focused on two aspects: the computational time required for feature extraction (if needed) and the computational time required for the ANN to make an inference. A detailed understanding of the computational demands for extracting features and running the model was thus obtained by measuring these parameters separately.

The performance of the deployed models was compared using several indicators: inference time, feature extraction time, total execution time, memory usage, and energy consumption. These factors are crucial in AE signal classification for applications like structural health monitoring, where quick processing and energy efficiency are essential for long battery life and immediate response. Energy consumption is estimated based on datasheet values of the processor's power consumption.

\section{Results and Discussion}\label{sec:results}

The presentation and discussion of results is separated into pre-deployment model evaluation, as well as results of the deployed models.

\subsection{Model Evaluation}

This section provides a detailed analysis of the ANN models' performance prior to their deployment. The models were assessed based on their classification performance based on accuracy, as well as their complexity in terms of parameter count and memory requirements. A summary of these results is given in Table \ref{tab:model_results}. Moreover, confusion matrices for each model are provided in Fig. \ref{fig:confusion_matrices}

\begin{table}[t]
    \centering
    \caption{Model test results}
    \begin{tabular}{lccc}
        \toprule
        \textbf{Model} & \textbf{Parameter} & \textbf{Accuracy} & \textbf{Model size} \\
        \midrule
        Raw-signal & 66\,243 & 0.996 & 258.76\,KB \\
        6-feature & 6\,979 & 0.992 & 27.26\,KB \\
        8-feature & 9\,283 & 0.991 & 36.26\,KB \\
        \bottomrule
    \end{tabular}
    \label{tab:model_results}
\end{table}

\begin{table*}
    \centering
    \caption{Key parameters of deployed ANN models}
    \begin{tabular}{lcccccc}
        \toprule
        \textbf{Model} & \textbf{Inference ($\mu$s)} & \textbf{Feature extraction ($\mu$s)} & \textbf{Total time ($\mu$s)} & \textbf{Flash size (KB)} & \textbf{RAM size (KB)} & \textbf{Energy (mJ)} \\
        \midrule
        Raw-signal & 6\,656 & -- & 6\,656 & 222.99 & 126.06 & 0.073 \\
        6-feature & 767 & 471\,937 & 472\,704 & 176.93 & 64.93 & 5.19 \\
        8-feature & 930 & 344\,106 & 345\,036 & 172.11 & 67.34 & 3.79 \\ 
        \bottomrule
    \end{tabular}
    \label{tab:deployment_results}
\end{table*}

Using the raw AE signals as input, the first model comprises 66\,243 parameters and thus is the most complex model among the three. This is to be expected, as each individual sample of the signal needs to be processed within the ANN model's neurons. It achieves the highest accuracy of 0.996, providing an excellent level of agreement between the predicted and actual classifications of AE signals. This high level of performance underscores the model’s capability to capture the intricate waveform patterns within the full spectrum of data points, enabling a robust classification of AE signals. Figure \ref{fig:confusion_matrices}a provides more details about the classification performance, illustrating a dominant number of true positives across all classes with very few misclassifications. 

The  6-feature model has significantly fewer parameters, totaling 6\,979. This reduction is due to the much smaller input size, reducing the number of parameters that need to be computed in each neuron. The accuracy is slightly lower than that of the raw-signal model, but with values of 0.992 the performance is still very high. Figure \ref{fig:confusion_matrices}b verifies this, demonstrating an overall low number of misclassifications without clear bias towards certain classes. Due to the much lower number of parameters, the model size has been significantly reduced to 27.26\,KB, enabling deployment on much more constrained embedded platforms.

The 8-feature model  excludes any frequency-domain features. Despite this limitation to time-domain features, the model achieves a similar classification performance as the previous model, with an accuracy of 0.991. It comprises 9\,283 parameters and is thus significantly larger than the model using the smaller feature set, but a low complexity in comparison to the raw-signal model. This results in a model size of 36.26\,KB, which is still very small. The confusion matrix presented in Fig. \ref{fig:confusion_matrices}c shows a very similar performance to the 6-feature model. This result demonstrates that frequency-domain features are not essential for an accurate classification in the given application scenario, but can reduce model size.

Moreover, the results highlight the trade-off between model complexity and performance. While the raw-input model achieves the best accuracy, it did so with many parameters, translating to a large model size (258.76\,KB). The reduced models, with significantly fewer parameters and smaller sizes, still performed admirably, providing lightweight alternatives based on careful feature selection. However, these models require feature extraction, which is analyzed in the next section.

\subsection{Deployment Results}

All three ANN models, as well as the required feature extraction algorithms have been deployed on an Arduino Nano 33 BLE Sense, encompassing the nRF52840 microcontroller. The deployed models are compared based on the inference time, feature extraction time, total processing time, model size, and energy consumption, all of which are listed in Table \ref{tab:deployment_results}. These metrics are fundamental to understanding the models’ performance in a constrained environment such as a microcontroller, where resources are limited and efficiency is paramount.

Due to the large number of parameters and operations, the model with raw signal input exhibits a relatively long inference time of 6\,656\,$\mu$s. On the other hand, it does not require any feature extraction. Hence, the total processing time is equivalent to the model inference time. The model size, including all required libraries, is the largest with 222.99\,KB Flash and 126.06\,KB RAM. The energy consumption for a single inference is estimated at 0.073\,mJ.

The 6-feature model, has a much lower inference time of only 767\,$\mu$s, but the time to extract the necessary features is substantial with 471\,937\,$\mu$s. This results in a total processing time of 472\,704\,$\mu$s. The model size was reduced to 176.93\,KB Flash and 64.93\,KB RAM, but energy consumption increases to 5.19\,mJ. As depicted in Fig. \ref{fig:model_comparison}, the substantial feature extraction time determines the total processing time.

\begin{figure}[t]
    \centering
    \includegraphics[width=.9\columnwidth]{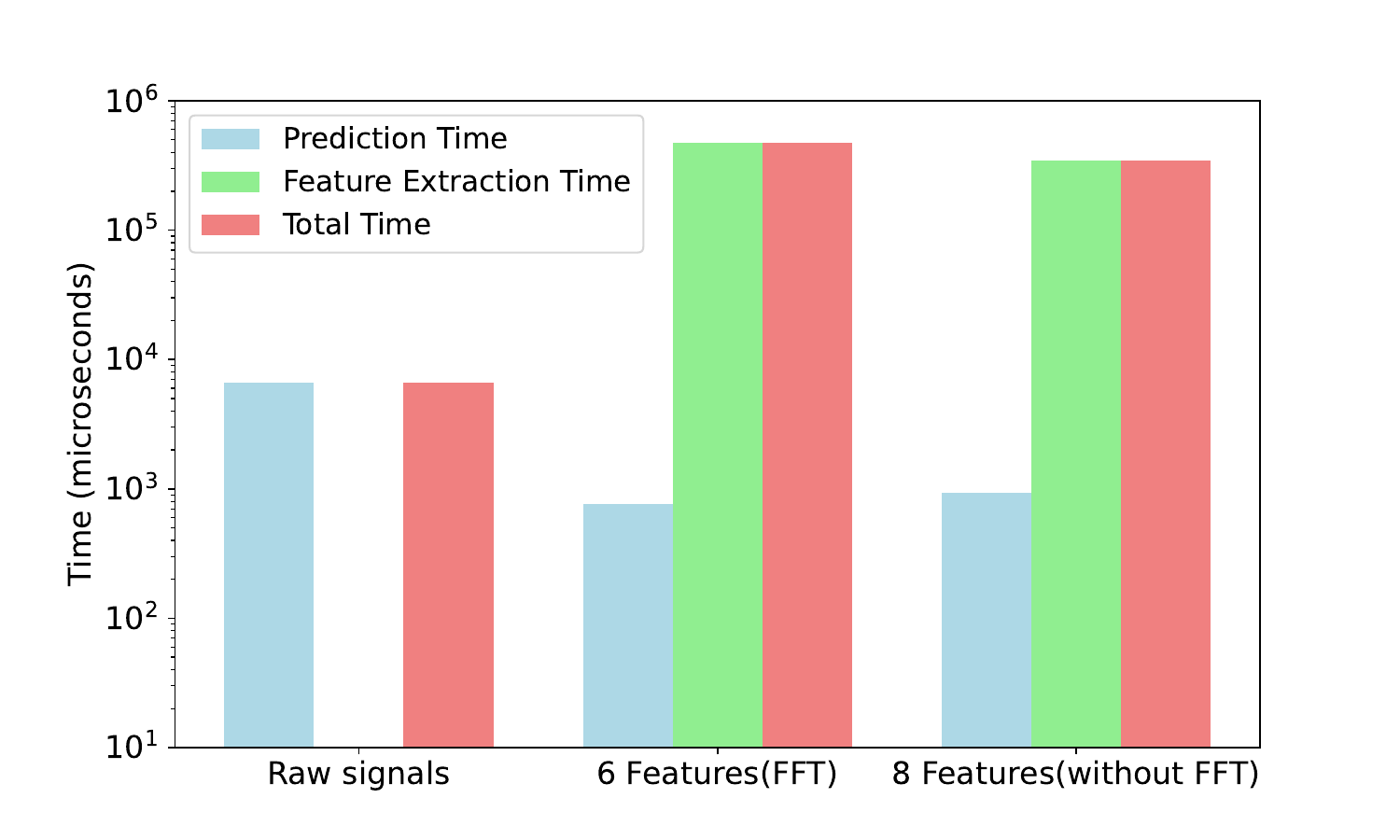}
    \caption{Comparison of inference time, feature extraction time, and total processing time for the three evaluated approaches.}
    \label{fig:model_comparison}
\end{figure}

Similarly, the 8-feature model, excluding any frequency-domain features, has a low inference time of 930\,$\mu$s, but a high feature extraction time of 344\,106\,$\mu$s. This solution thus amounts to a total processing time of 345\,036\,$\mu$s. The model size is about the same as the 6-feature model with 172.11\,KB Flash and 67.34\,KB RAM, while the energy consumption is estimated at 3.79\,mJ.

The comparison of the three approaches shows a clear trade-off between model complexity, processing time, and energy consumption. Despite its larger size and inference time, the raw-input model outperforms the other models in total prediction time and with minimal energy per prediction, making it the most suitable model for applications where the additional memory requirement is not a major concern. Conversely, while smaller in size, the 6-feature and 8-feature models require significant feature extraction times, which translates to high energy consumption. Due to the high feature extraction time, reduced model inference times become irrelevant as the total processing time is dominated by the feature extraction. Comparing these two models in more detail shows that removing frequency-domain features significantly reduces the feature extraction time (i.e., a reduction of about 120\,ms). Therefore, even though the 8-feature model is larger and has a higher inference time, its total processing time and energy consumption is significantly lower than those of the 6-feature model.

\section{Conclusions}\label{sec:conclusions}

This paper evaluated different TinyML approaches for the classification of AE signals with the aim to compare the use of extracted features with that of raw AE signals. Based on the presented results, several novel conclusions can be made. Firstly, despite its larger size, the raw-signal model demonstrated remarkable efficiency in inference time and energy consumption. This efficiency underscores the model’s ability to make rapid predictions, essential for real-time AE-based SHM systems. The absence of feature extraction simplifies the deployment process and enhances the model’s total processing speed, making it highly suited for scenarios where fast signal classification is necessary. While this performance comes at the cost of a larger memory requirement (222.99\,KB Flash, 126.06\,KB RAM), it is still compatible with many modern low-power microcontroller resources.

Conversely, while being more compact, the models operating on extracted features incurred significant computational overhead during the feature extraction process. This computational overhead resulted in longer total processing times and thus higher energy consumption of 71x and 52x of the raw-signal model, respectively, which could be detrimental in time-critical and energy-sensitive applications. However, these models maintained high classification performances, suggesting that a carefully selected subset of features can sufficiently capture the information required for accurate AE signal classification. It can also be concluded that the 8-feature model outperforms the 6-feature model, despite requiring a larger ANN model. This is due to the reduction in feature extraction time, as FFT computation is not required.

The classification performance of AE signals directly impacts the reliability of the SHM systems, which can be employed in various industrial applications. Accurate classification enables the early detection of potential faults in materials, preventing catastrophic failures and reducing downtime. The models developed in this study can distinguish between different modes of material stress with high accuracy, which is a testament to their potential utility in real-world applications. The contributions of this work demonstrate that such a SHM system is feasible to be directly implemented in resource-limited IoT devices.

The future trajectory of this research will focus on the integration with a complete IoT node that combines AE sensing and signal conditioning with the signal classification approach proposed in this work, enabling in-situ verification.

\section*{Acknowledgement}
The authors acknowledge financial support by the Knowledge Foundation under grant number 20180170 (NIIT).

\bibliographystyle{./IEEEtran}
\bibliography{./manuscript_review}

\end{document}